\documentclass[jgrga]{agu2001}
 \usepackage{graphicx}
 \usepackage{epsfig}

\authorrunninghead{SHAVIV}

\titlerunninghead{Cosmic Rays and the Faint Sun}

\authoraddr{Nir J. Shaviv,
Racah Institute of Physics, Hebrew University of Jerusalem, Jerusalem, 91904, Israel.
(shaviv@phys.huji.ac.il) }
\begin{document}

%
%

%
\setkeys{Gin}{draft=false}

%
%

%
%

\title{Towards a Solution to the Early Faint Sun Paradox:   A Lower Cosmic Ray Flux from a Stronger Solar Wind}
%

%
%


\author{Nir J. Shaviv}
\affil{Racah Institute of Physics, Hebrew University of Jerusalem, Jerusalem, 91904, Israel}

\begin{abstract}

Standard solar models predict a solar luminosity that gradually increased by about 30\% over the past 4.5 billion years. Under the faint sun, Earth should have been frozen solid for most of its existence. Yet, running water is observed to have been present since very early in Earth's history. This enigma is known as the faint sun paradox. We show here that it can be partially resolved once we consider the cooling effect that cosmic rays are suspected to have on the global climate and that the younger sun must have had a stronger solar wind, such that it was more effective at stopping cosmic rays from reaching Earth. The paradox can then be completely resolved with the further contribution of modest greenhouse gas warming. When we add the cosmic ray flux modulation by a variable star formation rate in the Milky Way, we recover the long term glacial activity on Earth. As to the future, we find that the average global temperature will increase by typically $10^\circ$K in the coming 2 Gyr.
\end{abstract}

%
%

%

\begin{article}

\section{Introduction}

According to standard solar models, the solar luminosity increased from about 70\% of the present solar luminosity at 4.5 Gyr before present (BP) to its present value. If Earth were a black body, its temperature would have been $\sim25^\circ$K lower, enough to have kept large parts of it frozen until about 1-2 Gyr BP. Besides however the past Eon, and the Eon between 2 and 3 Gyr BP, it appears that glaciations were altogether absent from the global surface. This is the crux of the faint sun paradox [{\em Sagan \& Mullen}, 1972; {\em Pollack}, 1991; {\em Sagan \& Chyba}, 1997].

A common solution to this apparent paradox is that larger amounts of the greenhouse gas (GHG) CO$_2$ compensated for the cooler sun [{\em Kuhn \& Kasting}, 1983; {\em Kasting}, 1993]. However, some upper limits on the atmospheric partial pressure of CO$_2$ ($p$CO$_2$) suggest that it was at most modest. For example, {\em Rye et al.} [1995] find $p$CO$_2 {\lower .75ex \hbox{$\sim$} \llap{\raise .27ex \hbox{$<$}} } 10^{-1.4}$~bar between 2.2 and 2.7 Gyr BP, with indications that it could not have been much higher  earlier [{\em Sleep \& Zahnle}, 2001].  At these levels, the GHG warming could at most compensate for about half the reduction in the solar flux\footnote{This assumes that a change of 1W/m$^2$ in the global radiation budget corresponds to the same global temperature change  irrespective of whether it is due to extra solar flux in optical and near-IR or the blocking of far-IR due to a higher level of $p$CO$_2$. This assumption is reasonable but not necessary.}. Thus, it appears that  CO$_2$ could have been important for warming the early Earth, but not sufficient to resolve the faint sun paradox by itself. Note however that theoretical considerations do suggest that $p$CO$_2$ was higher in the early terrestrial atmosphere [{\em Kasting}, 1987], such that it is likely that it did contribute at least some GHG warming.

Because of these problems, it was suggested that other GHGs could have been important for warming the young Earth. In particular, it was suggested that small amounts of NH$_3$ could have supplied the required GHG warming [{Sagan \& Mullen}, 1972; {\em Sagan \& Chyba}, 1997]. Although not impossible, it is not easy to keep NH$_3$ from irreversibly photolyzing into H$_2$ and N$_2$. 

Another suggestion was that CH$_4$ played the major GHG warmer [{\em Pavlov et al.}, 2000]. This solution requires a long residency time of methane in the atmosphere, and probably dominance of methanogenic bacteria. Unfortunately, there are currently no reasonable bounds on the amounts of methane or ammonia in the early atmosphere. Thus, this type of solutions can neither be ruled out, nor proven at his point.

Other effects are more certain but probably  not large enough to solve the faint sun paradox. In particular, the young Earth rotated faster than today and it may have had less land mass. Employing a typical general circulation model  (GCM), it was found that removing the land mass altogether could increase the temperature by 4$^\circ$K, while a 14 hr day would further increase the temperature by $1.5^\circ$K [{\em Jenkins}, 1993].

Another solution to the faint sun paradox involves strong solar winds, such that the  earlier solar mass was higher. A  10\% more massive sun would have had a larger luminosity,  enough to more than compensate for the early faint sun and also explain the observed Lithium and Beryllium depletion [{\em Graedel et al.}, 1991]. However, solutions with a relatively {\em gradual} loss of mass from the sun, with a typical time scale of $\sim$Gyr (as opposed to a mass loss time scale of a few 100 Myr) are inconsistent with helioseismology, rendering this solution  less probable [{\em Guzik \& Cox}, 1995]. Moreover, a solar wind history can be reasonably reconstructed if the sun is compared to similar nearby stars at various ages. This gives a benign mass loss which is insufficient to resolve the paradox at any time of the terrestrial history [{\em Wood et al.}, 2002]. 

Nevertheless, even a modest mass loss could potentially have interesting implications. The stronger wind was more effective at lowering the galactic cosmic ray flux (CRF). This, as we shall soon see, can present yet another possibility for helping resolve the faint sun paradox.  It should however be stressed that  it is not unlikely that the true solution is a combination of several of the above factors.

Various empirical evidence [{\em Tinsley \& Deen}, 1991; {\em Svensmark}, 1998; {\em Egorova et al.}, 2000; {\em Marsh \& Svensmark}, 2000; {\em Todd \& Kniveton}, 2001;  {\em Shaviv}, 2002a,b; {\em Shaviv \& Veizer}, 2003], theoretical analyses [{\em Ney}, 1959; {\em Dickinson}, 1975; {\em Harrison \& Aplin}, 2001; {\em Yu}, 2002], and experimental results [{\em Harrison \& Aplin}, 2001; {\em Eichkorn et al.}, 2002] link cosmic rays (CRs) to climate on Earth. The apparent CRF/climate link could originate through an effect that atmospheric ionization, which is governed by the CRF, may have on cloud formation. The empirical evidence is based on several significant correlations between the CRF and climate on Earth over times scales ranging from days to geological time scales. The theoretical and experimental evidence describe various elements of an emerging physical picture relating atmospheric ionization, through the formation of condensation nuclei by charged ion clusters, to the formation of clouds. 

Even with the above evidence, this CRF/climate link is still a highly debated topic. This is primarily because a robust physical picture is still missing. For an objective review, the reader is encouraged to read {\em Carslaw et al.}~[2002]. 
It should be emphasized that the analysis which follows rests on the assumption that this link indeed exists---an issue which we will address again in list of caveats in \S\ref{sec:discussion}. Irrespective of the outcome of this debate, since there is now enough evidence to seriously consider the possibility of CRs as a variable affecting the climate, we should consider the various implications of this link while bearing in mind that the underlying basis deserves extensive study.

We begin in \S\ref{sec:estimate} with an estimate of the total warming effect that the strong early solar wind has had, and compare it to the cooling induced by the fainter sun. In \S\ref{sec:evolution}, we use the actual solar wind evolution (as deduced from nearby stars) combined with the variable star formation rate (SFR) in the Milky Way (MW) to calculate the evolution of the terrestrial temperature. We summarize our results in \S\ref{sec:results} and discuss them in \S\ref{sec:discussion}.

\section{Estimating the effect of a stronger solar wind}
\label{sec:estimate}

We assume henceforth that cosmic rays do have a cooling effect\footnote{This can be seen in two ways. First, radiation budget measurements indicate that low altitude clouds have a net cooling effect [{\em IPCC}, 1995]. Thus, the positive CRF/low altitude cloud cover correlation [{\em Svensmark}, 1998) implies that a higher CRF will have a net cooling effect under the suspected CRF/climate link. Second, the long term {\em positive} correlation between CRF variations and ice-age epochs suggests that there is a {\em negative} correlation between CRF and temperature [{\em Shaviv}, 2002a, {\em Shaviv \& Veizer}, 2003].}.
Under this assumption, the stronger wind of the young sun had a warming effect. This is because the stronger wind necessarily lowered the galactic CRF reaching the inner solar system.
 A lower CRF would in turn translate into  higher temperatures through the CRF/climate link. Thus, a stonger solar wind should have had a warming effect that acted to compensate the solar dimness.

To quantify this effect,  two basic numbers are required. First, it is necessary to know the ratio between changes in the CRF and changes in the global temperature. This will allow us to translate solar wind variations into temperature change on Earth. Second, if we wish to compare this effect with the lower temperature associated with the faint sun, and also with possible GHG warming, we need to know how changes in the radiation budget translate into a global temperature change.

The relation between radiative forcing and temperature change is still an open and highly discussed question in climatology. For a blackbody Earth, this number is 0.30$^\circ$K/(W/m$^2$). Earth however does not behave as a black body, and various positive and negative feedbacks act to either amplify or decrease this number by a factor $\lambda$. Values obtained in GCMs [{\em IPCC}, 1995] range between $\lambda=2$ and $4.5$. 

There are however debated claims that various negative feedbacks should be considered, yielding a much smaller $\lambda$ of $0.5-1.3$ [{\em Lindzen et al.}, 2001], or even $\sim 0.2$ [{\em Ou}, 2001]. This number was also obtained from analyses of actual temperature variations. For example, from the rather quick return of the global temperature to its ``average" after a series of volcanic erruptions, it was deduced that $\lambda \sim 0.5$  with $\lambda { \lower .75ex \hbox{$\sim$} \llap{\raise .27ex \hbox{$>$}} } 1$ being inconsistent with the effects of volcanic driving [{\em Lindzen \& Giannitsis}, 1998]. On the other hand, {\em Gregory et al.} [2002] find that a lower limit of $\lambda=1.3$ (at 95\% confidence) can be placed by considering the increase of ocean heat content. Another constraint was obtained by {\em Covey et al.} [1996] who compared $p$CO$_2$ changes to temperature variations over geological time scales (in particular, the last glacial maximum, the Crataceous and early Eocene). The authors of this study found that $\lambda$ is consistent with the range of values obtained in GCMs.  However, the latter two constraints were obtained while neglecting the possible forcing of the CRF and the nonthermal solar activity. Thus, if CRF/climate link is real, these two constraints are not applicable anymore.
 
Once CRF variations are considered over geological time scales, a markable correlation is obtained between CRF as a driver and temperature change. This correlation,  combined with the lack of correlation with $p$CO$_2$ variations, yields  $\lambda \sim 0.75$ (or $\lambda < 1.2$ at 90\% confidence and $\lambda < 2.2$ at 99\% confidence) [{\em Shaviv \& Veizer}, 2003]. Thus, {\em if the CRF/climate connection is valid}, which we assume here to be the case, we should consider $\lambda \sim 0.75$ as the most probable value, but also $\lambda = 2.2$ as an extreme upper limit.

The same combined geological and astronomical analysis [{\em Shaviv \& Veizer}, 2003]
could also measure the relation between temperature change and relative CRF variations, which was parameterized through the form 
\begin{equation}
\label{eq:CRFT}
\Delta T_{CRF} \approx D \left[ 1- ({\cal E}/{\cal E}_0)^q \right],
\end{equation}
where ${\cal E}$ is the energy flux reaching the troposphere (to which the ionization is proportional), and ${\cal E}_0$ is the flux reaching today.
The power $q$ relates changes in the ionization rate to changes in global temperature. Since the expected route is most likely through the effect that atmospheric ionization has on the formation of condensation nuclei, we should have $q\approx {1 \over 2}$  [{\em Yu}, 2002; {\em Harrison \& Aplin}, 2001]. It was also found that $D=10 \pm 5^\circ$C, where most of the uncertainty is due to the inaccurate determination of the CRF variations from meteorites. We reduced here the upper range. Although a larger $D$ is consistent with the meteoritic data, it is physically inconsistent as it requires too large an albedo variation of Earth to be physically reasonable, as will soon be evident.

When the solar system was young, the solar wind is expected to have been significantly stronger. This implies that most of the CRF responsible for tropospheric ionization was blocked (i.e., the energy flux satisfied ${\cal E} \ll {\cal E}_0$). Today, for comparison, the flux variations are only ${\cal{O}}(10\%)$. Thus, we should expect an average global surface temperature increase of $\Delta T_{wind} \approx D \approx 10 \pm 5^\circ$K, when the wind was significantly stronger than today. 

Next, the decrease in solar luminosity as we  go back  to the end of the heavy bombardment era (at about 4 Gyr BP), is about $\sim$25\%. This is an inevitable result of all standard models\footnote{It originates from the fact that as nuclear burning proceeds, the molecular weight of the solar core becomes progressively larger.}. The only astrophysical remedy is having a heavier younger sun. However, as mentioned in the introduction, a strong and slowly decaying wind that would allow the young sun to be bright for a long enough duration, is inconsistent with the winds of nearby stars [{\em Wood et al.}, 2002].

The $25\%$ reduction in solar flux corresponds to a decrease of  $\sim$60~W~m$^{-2}$ relative to the $240$~W~m$^{-2}$ reaching the surface today. For $\lambda = 0.5,0.75$ or $2$, the average global surface  temperature decrease $\Delta T_\odot$ associated with the fainter sun is $\Delta T_{\odot} = - 9^\circ$K, $-13.5^\circ$K or $-36^\circ$K respectively. Thus, for climate sensitivities consistent with the CRF/climate link ($\lambda {\lower .75ex \hbox{$\sim$} \llap{\raise .27ex \hbox{$<$}} } 0.75$), we find that  the solar dimness can in principle  be completely compensated for. 

Nevertheless, to actually obtain $\Delta T_{wind} > \left|\Delta T_{\odot}\right|$, we require values of  $\lambda$ and $D$ which are consistent with observations but are theoretically uncomfortable. For example, $D=15^\circ$K with $\lambda = 0.5$ implies that CRs change the albedo by 28\%, which is nearly impossible.    The more reasonable values of $D=10^\circ$K and $\lambda=0.75$ yield a more realistic albedo change of 12.5\%. In this case  however, $\Delta T_{wind} + \Delta T_{\odot} \approx -3.5^\circ$K relative to today. This implies that the average Hadean and Archean temperatures would have been well above freezing. However, a full resolution of the faint sun paradox requires the temperatures at these epochs to have been higher than today, since current day Earth  has glaciations, while glaciations where absent from the young Earth (e.g., {\em Crowell} [1999]).

With $\lambda = 0.75$, the missing $3.5^\circ$K can be compensated with $16$~W~m$^{-2}$. This could arise from  a ``modest" GHG warming.  For example, $0.02$~bar of CO$_2$, which is  about half of the present upper limits at $\sim 3$ Gyr BP [{\em Rye et al.}, 1995; {\em Sleep \& Zahnle},  2001] would provide $\sim 32$~W~m$^{-2}$ of GHG heating\footnote{We assume that the radiative forcing of $CO_2$ on the atmosphere is $F_{CO2} = 4 W/m^2 \left[1.236 \ln(c+0.005 c^2)-(\ln(c_0+0.005 c_0^2)\right]$ where $c_0=280 ppm$ is the pre-industrial level of atmospheric CO$_2$ [{\em Hansen et al.}, 1998]}, ensuring that the early climate was warm enough to inhibit the appearance of glaciations.

\section{Solar wind evolution and predicted Temperature change}
\label{sec:evolution}

To estimate the secular change in the solar wind modulation of the CRF, we first require the evolution of the solar wind parameters (i.e., mass loss rate and velocity). We will then  translate this evolution to variations in the CRF reaching Earth.

For the evolution of the mass loss rate $\dot{M}$, we use the observational results of {\em Wood et al.} [2002] who found that the mass loss rate of nearby solar-like stars (main sequence stars of spectral type G and K) evolves as: $\dot{M} \propto t^\mu$ with $\mu=-2.00 \pm 0.52$. 
Next, we take the velocity of the wind to be constant. This is because the terminal wind velocity depends primarily on the escape velocity from the sonic radius of the wind, which is expected to change only logarithmically with the mass loss rate (e.g., {\em Lamers \& Cassinelli} [1999]). This implies that to first approximation we can take $\delta \approx 0$ in a relation of the form $v_{wind} \propto \dot{M}^\delta$, but we leave $\delta$ as a free parameter for generality. 

The energy loss suffered by CRs as they journey from the heliopause to Earth is through multiple scattering off the magnetic field of the solar wind. The average energy lost, for CRs with energies significantly larger than their rest mass, can be approximated with $\Delta E \approx R v_{wind} /(3 A)$ [{\em Perko}, 1987], where $R$ is the radius of the heliopause and $A$ is a constant proportional to $\kappa$, the diffusion coefficient of CRs in the solar wind. This expression is valid {\em only if}  $\Delta E \ll E$, $E \gg m_0 c^2$ and $R \gg 1 AU$. Thus, for the calculation of the energy loss we further require the scaling of both $R$ and $\kappa$ with $\dot{M}$.

The value of $R$ can be obtained by equilibrating the solar wind ram pressure at its termination $\dot{M}v_{wind}/R^2$ with the average pressure of the interstellar medium (ISM), which we assume to remain constant on average. This yields $R \propto \sqrt{\dot{M}v_{wind}}\propto \dot{M}^{(1+\delta)/2}$.

The diffusion coefficient $\kappa$ depends on the effective mean free path $\ell$ arising from the Larmour gyration of the CRs in the disordered magnetic field of the solar wind. Since $\ell \propto B^{-1}$, we have $\kappa \sim \ell c \propto B^{-1}$. Thus, the last scaling we seek is in the form $B \propto \dot{M}^\gamma$. It would allow us to relate $\dot{M}$ to changes in $\Delta E$.

At the radius of Earth's orbit, the magnetic field energy density in the solar wind is typically 50 times smaller than the kinetic energy density of the wind. However, magnetic field fluctuations should decay as $r^{-3/2}$ [{\em Jokipii},1973], (whether frozen-in or Alfv\'enic wave fluctuations), implying that at a few solar radii, where the solar wind is accelerated, the magnetic field energy density ($\propto r^{-3}$) is comparable to the kinetic energy density ($\propto r^{-2}$). Although the current mass loss is probably not unique, we find equipartition between the two energy densities at the acceleration region of the wind. Thus, it is reasonable to assume that this tendency for equipartition remains also for different $\dot{M}$'s.  In such a case, $B^2 r^{3/2} \propto \rho v_{wind}^2 \propto \dot{M} v_{wind}$ at any given radius. This implies that $\gamma =(1+\delta)/2$ if equipartition remains.

The combination of these scalings yields $\Delta E \approx \dot{M}^{1/2+\gamma + 3\delta/2} \propto t^{\tilde{\mu}}$, with $\tilde{\mu} \equiv \mu(1/2+\gamma + 3\delta/2)$. With nominal mass loss evolution ($\mu \approx 2$), constant wind velocity ($\delta\approx 0$)  and aforementioned equipartition ($\gamma\approx 1/2$),  we have $\tilde{\mu} \approx \mu \approx -2$.

The CRF spectrum (i.e., the differential number flux) in the ISM can be written as $d\Phi/dE_{ISM} \equiv f_{ISM} = {\cal C}E^{-(p+1)}$ for energies greater than about $1$~GeV, where $\Phi$ is the CR number flux, ${\cal C}$ is a normalization constant,  and $p\approx 1.7$. Furthermore, the differential number flux $f_{\oplus}$ reaching Earth is related to $f_{ISM}$ through $f_{\oplus}(E) = f_{ISM}(E+\Delta E)$. Since only CRs with an energy larger than  
the cutoff energy $E_c \sim 12$~GeV can actually reach the troposphere, we find that the total energy ${\cal E}_{>E_c}$ deposited in the troposphere is 
\begin{eqnarray}
{\cal E}_{>E_c} &=& \int_{E_c}^{\infty} f_{\oplus} E_{\oplus} dE_\oplus \\ \nonumber &=& 
\int_{E_c}^{\infty} {\cal C} (E_{\oplus}+\Delta E)^{-(p+1)} E_\oplus dE_\oplus\\&=&  {\cal C}(E_c+\Delta E)^{-p}\left({E_c\over p-1} + {\Delta E \over p}  \right). \nonumber
\end{eqnarray}
More interesting is the energy flux at a time $t$ (after the formation of the solar system) relative to the flux today, at $t_0$. We define $e\equiv \Delta E_0 / E_c$. Since $\Delta E_0 \approx 1.0$~GeV [{\em Perko}, 1987], we have $e \approx 0.08$, and also
\begin{equation}
{{\cal E}_{>E_c}(t) \over {\cal E}_{>E_c}(t_0)}= { \left( p + \left(t / t_0\right)^{-\tilde{\mu}}e \over  p +e  \right) }{\left(1+\left(t / t_0\right)^{-\tilde{\mu}}e \over 1+e \right)^{-p}}.
\end{equation}

Eq.~\ref{eq:CRFT} for the temperature change induced by the wind evolution now becomes:
\begin{equation}
\Delta T_{wind} \approx \left[ 1- \left({{\cal E}_{>E_c}(t) \over {\cal E}_{>E_c}(t_0)}  {{\cal S}(t) \over {\cal S}(t_0)}\right)^q\right] D,
\end{equation}
where ${\cal S}$ is the nearby SFR to which the CRF is proportional\footnote{Almost all supernovae in the MW are of  types other than Ia. These originate from the death of massive stars which live less than $\sim 30$ Myr. Because this time scale is practically instantaneous (relative to 4 Gyrs) and because CRs (at 10 GeV) originate from supernova explosions, the CRF will be proportional to the SFR (see references in {\em Shaviv} [2002b]).}.
For the SFR, we take the results of {\em Rocha Pinto et al.} [2000] who derived the MW SFR by analyzing the chromospheric ages of hundreds of late type dwarfs. The results of this comprehensive study are consistent with previous  analyses, such as {\em Barry} [1988]  who too used chromospheric ages, or {\em Scalo} [1987] who studied the mass distribution of nearby stars. In particular, it appears that the SFR was higher than average between 2 and 3 Gyr before present, while it was below average between 1 and 2 Gyr BP.

 Interestingly, the MW SFR activity is correlated with SFR activity
 in the Large Magellanic Cloud (LMC). It appears that there was a significant increase in the SFR in the LMC around  2 Gyr BP [{\em Gallagher et al.}, 1996; {\em Vallenari et al.}, 1996; {\em Dopita et al.}, 1997] and that 
 between 0.7 to 2 Gyr BP, the SFR in the LMC was below its average [{\em Westerlund},1990].  
 This correlation could be coincidental, but may reflect the tidal interactions between the two galaxies, during their perigalactica sometime around 0.2-0.5 Gyr BP and around  1.6-2.6 Gyr BP [{\em Gardiner et al.}, 1994; {\em Lin et al.}, 1995]. Irrespective of the origin of the SFR variability, it modulated the CRF reaching the solar system, and presumably also climate on Earth.

The fact that  the moon was closer to Earth, and both were rotating faster [{\em Zharkov}, 2000] and that Earth probably had a smaller land mass are not large effects by themselves [{\em Jenkins}, 1993]. We will not try to model them accurately. Nevertheless, {\em Jenkins} [1993] has shown that no land mass and a 14 hr day will correspond to an average global temperature increase of typically 5.5$^\circ$K. This however was obtained with GCM models which typically have $\lambda \sim 2.5$. If, on the other hand, negative feedbacks exist to reduce $\lambda$, the increase in temperature will be smaller accordingly. For $\lambda \approx 0.75$, we should expect that the
whole effect of no land mass and a faster rotation would contribute only a $\sim 1.6^\circ$K warming that gradually decreases with time between the formation of Earth and today.

 To mimic these two effects, we approximate the temperature correction using $\Delta T_{corr}(t) = \lambda \left[(t_0-t)/t_0 \right]  2.2^\circ$K where $t_0=4.5$~Gyr is the time today. This is not a large correction at all, but unlike effects such as GHG warming by NH$_3$ or CH$_4$, which can potentially be much larger, the effects of land mass and rotation can be considered  somewhat more quantitatively.





\section{Results}
\label{sec:results}

Using the above physical components, we proceed to calculate the evolution of the average terrestrial temperature. The results are depicted in fig.~1 for various models which either include or exclude the solar wind/CRF effect using nominal values for the parameters,  the effects of radiative forcing, and also the possibility of  GHG warming by modest levels of $p$CO$_2$.

\begin{figure*}[htb]
\centerline{\includegraphics[width=28pc]{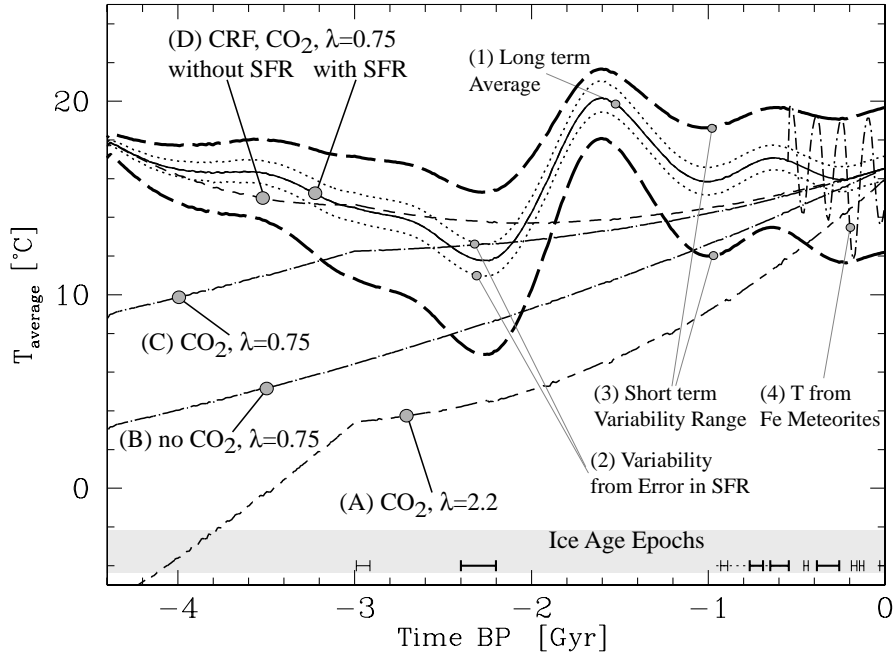}}
\caption{ 
Predicted temperature as a function of time before present for various models. Model (A) includes a radiative feedback of $\lambda=2.2$ and the GHG warming by modest levels of CO$_2$ (by 0.01 bar of CO$_2$ before 3 Gyr and exponentially decreasing afterwards to current levels). Model (B) includes no CO$_2$ contribution but $\lambda=0.75$. Model (C) has both CO$_2$ and $\lambda=0.75$, but still has no CRF contribution. Model (D) includes CRF contribution using nominal values ($\tilde{\mu}=2$, $E_c=12$~GeV, $\Delta E_0=1.0$~GeV, $D=10^{\circ}$C). The short dashed line assumes the CRF reaching the outskirts of the solar system is constant and equal to the current day average. The solid line (no.1) assumes that the CRF is proportional to the SFR in the Milky Way [{\em RochaPinto et al.}, 2000]. The dotted lines (no.2) depict the temperature uncertainly associated with the SFR uncertainly error. The temperature variability associated with ``short term" variability in the CRF, originating from the spiral structure [{\em Shaviv}, 2002a,b], is marked with the long dashed lines (no.3). These denote the temperature variability range. The actual temperature variability in the past 500 Myr [{\em Shaviv \& Veizer}, 2003], which is associated with the variable CRF (as extracted from Iron meteorites [{\em Shaviv}, 2002b]), is marked with the short-dash-dotted line (no.4).}
\end{figure*}


The figure recapitulates the main conclusion of \S\ref{sec:estimate}. Namely, coupling the CRF/climate effect to the stronger wind of the young sun, results with a higher temperature. The effect is large, but if the nominal values are used it compensates only $\sim {2\over 3}$'s of the temperature reduction associated with the dim sun. Thus, either the CRF effect is larger than inferred from the nominal parameters, or more likely, there is some GHG warming.

By estimating the actual evolution of the temperature, we find additional conclusions to those obtained in \S\ref{sec:estimate}. First, the effects of SFR variability, spiral arm passages and GHG warming are all comparable in amplitude. Second, because of the different time scales associated with the slowly varying SFR and the ``fast" oscillations associated with spiral arm passages (over $\sim 10^8$~yr), we find that at any epoch with a given SFR, a range of temperatures should exist from spiral arm variability. In other words, in addition to the statistical error on the average SFR [{\em RochaPinto et al.}, 2000], we should consider that the location within the MW gives rise to additional variability\footnote{In the past Eon, for example, the average CRF was about $\sim 70\%$ of the current CRF [{\em Lavielle et al.}, 1999; {\em Shaviv}, 2002b] but periods during which the solar system spent near spiral arms, the solar system was bathed with a flux higher than today, which presumably reduced the global temperatures enough to allow glaciations.}. Thus, not only is the {\em average} important, but also the expected lower limit of the temperature range if we wish to compare with geological evidence for the presence of glaciations. On the other hand, if we had a good record for the presence of extremely high temperatures, then the expected upper limit of the temperature range would have been the important variable. 

If we study fig.~1 and consider models with the nominal parameters of $\lambda \approx 0.75$ and $D\approx 10^\circ$K, we find that only during the past Eon (Neo-Proterozoic--Phanerozoic period), and the period of  2-3 Gyr BP (late-Archean--Huronian)  could there have been climatic periods during which it was cold enough as to produce temperatures lower than today and explain glaciations. This is consistent with the actual geological evidence, which indicates that only during the Neo-Proterozoic and Phanerozoic (past Eon), during the Huronean (2.2-2.4 Gyr BP) and late Archean (2.9-3.0 Gyr BP) were glaciations present on Earth (e.g., {\em Crowell} [1999]). 

The exception is the first eon, which for the nominal parameters should have been cold enough to have had glaciations (though not as cold as 2.5 Gyr BP). Since there is no evidence to support their existence, we return back to the point that even a modest GHG contribution, for example,  by a modest 0.01~bar CO$_2$ atmosphere, can raise the temperature of the very young Earth such that with the reduced CRF, the temperature is kept above today's values. 

\begin{figure}
\centerline{\includegraphics[width=18pc]{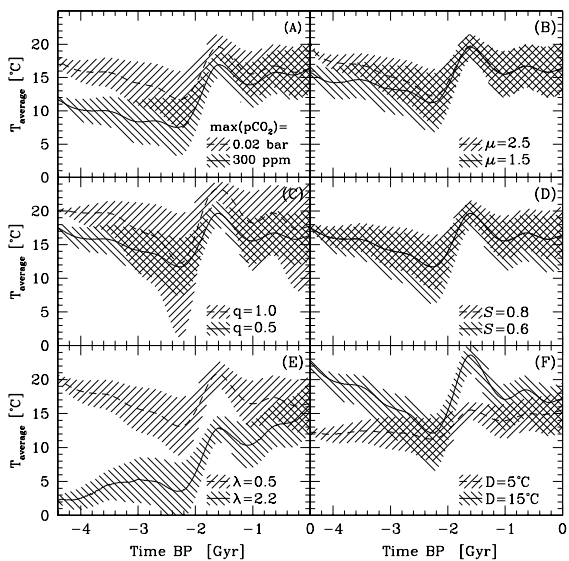}}
\caption{ 
A parameter study. Each panel describes models in which one parameter is changed from the nominal model of fig.~1. The average global temperature is depicted here with either a solid or dashed line (equivalent to the solid line of fig.~1). The expected range of variability
arising from spiral arm passages is depicted here with hatched shading (equivalent to the two-long dashed lines of fig.~1).   {\em Panel (A)} describes  the effect of $p$CO$_2$. The solid line is the nominal model with $p$CO$_2$ kept at todays value while the dashed line has a saturation $p$CO$_2$ of 0.02 bar before 3 Gyr BP. {\em Panel (B)} describes the effect of changing the mass loss evolution power law $\mu$ (within observed error range). {\em Panel (C)} describes the effect of the power $q$ which relates the atmospheric ionization rate change to temperature change. {\em Panel (D)} describes the effect of changing the ratio between the average SFR in the past few 100 Myr and the current rate (nominal value is ${\cal S}=0.7$, {\em Lavielle} [1999]). {\em Panel (E)} describes the effect of a changed climatic sensitivity to radiation forcing. {\em Panel (F)} describes the effect of a changed climatic sensitivity to CRF changes. The normalization is such that for today's flux, which is about $30\%$ higher than the average over the past eon, the temperature is $15^\circ$C.
}
\end{figure}


Fig.~2 is a parameter study showing the effects of changing various model parameters from their nominal values. It is apparent from the figure that there are two types of parameters. The first type are those which affect the total CRF/solar wind effect  when the solar wind is completely switched off, or affect the way in which the effect of the dim sun translates into a temperature change. These parameters determine the gross temperature behavior. For example, $\lambda$, $D$ and maximum level of $p$CO$_2$ determine the temperature on the very young Earth. The second type of parameters are secondary in the sense that they only affect the evolution of the temperature between the young Earth and today.  A good example is the power  $\mu$ determining the temporal evolution of $\dot{M}$. It does not affect the temperature in the extremely young atmosphere nor the temperature today, since the former is obtained in the limit of a very strong wind (irrespective of how strong it actually is), and the latter is simply given by the current $\dot{M}$. 

Among the secondary parameters, the most important is the power $q$ which relates the ionization rate to temperature fluctuations. The value $q={1 \over 2}$ is theoretically preferred [{\em Harrison \& Aplin}, 2001; {\em Yu}, 2002]. A higher $q$ would imply that CRF variability arising from a variable SFR and passages through spiral arms becomes more prominent. Nevertheless, all the secondary parameters are not particularly important.

\begin{figure*}
\centerline{\noindent\includegraphics[width=30pc]{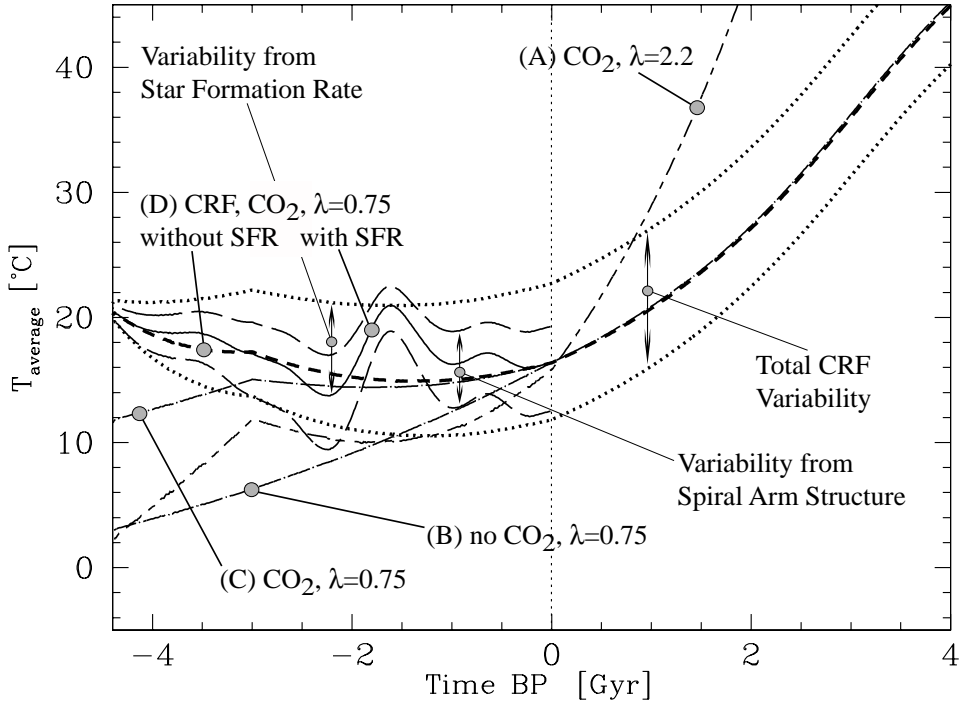}}
\caption{ 
Past and future long term global temperature changes. The graphs depicted in this figure are the same as fig.~1, with a few exceptions. (a) ``Short term" ($\sim 10^8$yr) temperature variability is not plotted. (b) Models with CO$_2$ have $p$CO$_2=0.04$~atm before 3 Gyr BP (consistent with upper limit by {\em Rye et al.} [1995]), then decrease exponentially until today and remain at today's levels in the future. (c) The future temperature trend is plotted. Since the future SFR is unknown, the upper and lower temperatures associated with the actual SFR and spiral structure cannot be extended into the future. Instead, only the total range associated with both past SFR  and spiral variability can be plotted. In the first $\sim 5$ Gyr of Earth's existence, the temperature increase due the increased luminosity is compensated for by the weaker solar wind and ensuing higher CRF. Since almost all the CRF (causing tropospheric ionization) currently reaches the inner solar system, this compensation cannot continue and the temperature will start to significantly increase.     
}
\end{figure*}

Although future variability of the MW SFR cannot be predicted, we can predict the future evolution of the solar luminosity and the decrease in strength of the solar wind. Since the solar wind currently filters out only about 10\% of the cosmic rays responsible for tropospheric ionization (with energies ${ \lower .75ex \hbox{$\sim$} \llap{\raise .27ex \hbox{$>$}} } 10$~GeV), the cooling  effect of the  solar-wind/CRF coupling is almost at its fullest. This implies that from now on, in the coming few billion years, the global temperature is expected only to increase on average. This is exemplified in the fig.~3. This is not to say that bursts of star formation could reduce the temperature  ``temporarily", for example, when the LMC will be swallowed by the MW.

We expect that within  about 1.5 Gyr, the long term average global temperature will increase to values higher than anytime previously (i.e., ${ \lower .75ex \hbox{$\sim$} \llap{\raise .27ex \hbox{$>$}} } 20^\circ$C), at which point the negative temperature excursions arising from spiral arm passages and a higher SFR will not cool Earth down to current day temperatures, and no more ice-age epochs will ever appear. From about 2 Gyr into the future, even spiral arm passages and a higher SFR will not be able to reduce the global temperature to below the warmest temperature experienced in the past 4 Gyr.

\section{Summary}
\label{sec:discussion}
We combined here the proposition that CRF variations affect the global climate with the prediction that the early solar wind should have been stronger than today. This was shown to result  with the following conclusions:
\begin{enumerate}
\item The stronger wind of the young sun was more effective at blocking the CRF reaching Earth such that the CRF was significantly reduced when compared with the current flux.
\item When coupled to the suggested CRF/temperature link, the lower CRF  reaching Earth should have translated into a higher global temperature.  
\item Using the apparent correlation between CRF variability and global temperature     over the past 550 Myr [{\em Shaviv \& Veizer}, 2003], it is possible to quantify this effect. It was found that the associated warming was enough to significantly compensate for the {\em faint sun} and explain about $1\over 2$ to $2 \over 3$'s of the temperature increase required to warm the young Earth to above the average temperature today. The remaining $1 \over 3$ to $1 \over 2$ could come from one of the previously suggested factors, such as GHG warming by {\em modest levels} of CO$_2$ ($p$CO$_2$ of order 0.01 bar), which are consistent with the geological constraints.
\item  Besides explaining the secular climate trend, the evolution of the solar wind  also reproduces, essentially without any free parameters, long term temperature trends on Earth once the variable SFR in the MW is considered. In particular, the CRF evolution explains why Earth has had glaciations in the past Eon and between 2 and 3 Gyr BP. 
\item  Although the ``fine" details of a variable SFR and the passage through spiral arms cannot be predicted to more than several 100 Myr into the future, it is possible to extrapolate the decrease in solar wind and the increase in solar flux. It is found that from about 1.5 billion years into the future, Earth is expected to be warmer  on average  than anytime in its previous history. This is because the CRF reaching Earth is currently about as large as it can be.  
\end{enumerate}

These conclusions, however, do not come without {\em caveats} that are at least as important as the conclusions themselves.

\begin{enumerate}
 \item First and foremost, the aforementioned conclusions rest on the assumption that the CRF/climate connection is indeed real. Since this link still lacks a robust physical picture to support it (and is therefore still a subject of debate), it clearly deserves extensive study,  especially in view of the repercussions that it has. For example, it  bears consequences on the value of the radiative forcing, since only a low value is consistent with the CRF/climate link.
 \item Even if we were safe to assume that the CRF climate connection is real, since the process is not fully understood, it is not trivial to guaranty that it also operated before the past Eon, during which periodic ice-age epochs existed. In particular, the CRF/climate link could depend on a biogenic agent which may have been absent during the Archean or early Proterozoic Earth. The link could also depend on atmospheric composition which undergone a few notable changes over the Eons.
 \item We have neglected in the present analysis additional effects that may be important but hard to quantify. These include for example effects of atmospheric composition, land mass characteristics (area and albedo), and even galactic migration of the solar system. 
\end{enumerate}


\begin{acknowledgments}
This research was supported by the F.I.R.S.T. (Bikura) program of the Israel Science Foundation (grant no. 4048/03).
\end{acknowledgments}

\end{article}
\newpage

\end{document}